\documentclass{svjour2}                    
\smartqed  
\usepackage{graphicx}
\usepackage{color}
\newcommand {\be} {\begin{equation}}
\newcommand {\ee} {\end{equation}}
\newcommand {\bq} {\begin{eqnarray}}
\newcommand {\eq} {\end{eqnarray}}

\begin{document}

\title{Portable implementation of a quantum thermal bath for molecular dynamics simulations
}


\author{Jean-Louis Barrat         \and David Rodney
}


\institute{Jean-Louis Barrat \at
              Universit\'e Grenoble 1 and CNRS, Laboratory for Interdisciplinary Physics,
               UMR 5588, Saint Martin d'H\`eres, F-38402, France
              \email{jean-louis.barrat@ujf-grenoble.fr}           
           \and
            David Rodney \at
      SIMaP, Grenoble INP, UJF, CNRS, UMR 5266, Saint Martin d'H\`eres, F-38401 France
                                       \email{david.rodney@grenoble-inp.fr}           
}

\date{Received: date / Accepted: date}

\maketitle

\begin{abstract}
Recently, Dammak and coworkers \cite{dammak-prl2009} proposed
 that the quantum statistics of vibrations in condensed systems at low temperature could be simulated by running
molecular dynamics simulations in the presence of a colored noise with an appropriate power spectral density.
In the present contribution, we show how this method can be implemented in a flexible manner and at a low computational cost
by synthesizing the corresponding noise 'on the fly'. The proposed algorithm   is tested for a simple
harmonic chain as well as for a more realistic model of aluminium crystal. The energy and Debye-Waller factor are shown to be
in good agreement with those obtained from harmonic approximations based on the phonon spectrum of the systems. The limitations
of the method associated with anharmonic effects are also briefly discussed.
Some perspectives for disordered materials and heat transfer are considered.
\end{abstract}

\section{Introduction}
\label{intro}

Standard constant-energy molecular dynamics (MD) simulations explore the phase space of a system by following the classical equations of motion, therefore producing a classical micro canonical distribution function. Very early it was realized that these equations could be modified in several different manners to produce a canonical phase space distribution at fixed temperature. Among the possible ways to achieve this is the implementation of a ''Langevin thermostat'' \cite{kremer1986} that couples each particle
 independently to a fictitious thermal bath by introducing in the equation of motion a random force and a friction term related by the classical fluctuation-dissipation theorem. This is illustrated by the following equation for a single degree of freedom $X$ of mass $M$ submitted to an external force $F(X)$:
 \begin{equation}
 M \frac{dV}{dt} = - M \gamma V + {F}(X) + \sqrt{2 M \gamma}~\theta(t)
 \label{langevin}
\end{equation}
where $V=dX/dt$ is the velocity, and $\theta$ is a white noise verifying:
 \begin{equation}
 \langle \theta(t)\theta(t')\rangle = k_BT \delta(t-t').
\end{equation}
While this type of thermostat (and various other more sophisticated versions developed by Hoover, Nos\'e and others, see for example Ref. \cite{FrenkelSmit}) provides a convenient control of the temperature, it is intrinsically classical, and in particular the equipartition theorem
is strictly verified. Any quadratic degree of freedom will on average correspond to an energy $k_BT/2$. While this is a minor problem in
liquid systems at room temperature, in which quantum effects are effectively rather small, the situation is quite different in solids. The Debye temperature, that signals the onset of quantum effects, is often of the order of a few hundred Kelvins, meaning that quantum effects are indeed important. The most prominent manifestation of this fact is the deviation from the Dulong and Petit law of heat capacity, which is
manifest in most solids already at room temperature.

Obviously quantum dynamics in solids is well handled by the usual harmonic theory of solids, possibly including anharmonic corrections. Using this theory however requires a full calculation of the normal mode spectrum, which may be cumbersome for large or aperiodic systems. Calculations of transport properties such as heat conductivity also represents a real challenge, as well as static properties in strongly anharmonic systems. While static
aspects can be dealt with using path integral methods, dynamical effects can not. Moreover, path integrals represent at low temperatures a rather costly alternative to much simpler classical simulations due to the large number of replicas to include in the path integrals in order to converge to the quantum limit \cite{muser_jcp2001}.

It is therefore highly desirable to develop methods that would allow one to describe
correctly quantum effects at a relatively low computational cost, even if the representation is approximate. Recently two very interesting steps in this direction were taken by  Dammak \textit{et al} \cite{dammak-prl2009} and by Ceriotti  \textit{et al} \cite{ceriotti-prl2009}. The fundamental idea introduced in both papers is that the correct distribution of positions and impulsions for a \emph{quantum} harmonic oscillator can be reproduced by using an equation similar to Eq. \ref{langevin} but with a colored noise replacing the white noise $\theta(t)$.
In Ref. \cite{ceriotti-prl2009}, the friction term was moreover substituted by a Non-Markovian term, thus permitting a greater flexibility in the
types of thermostats used. This non-Markovian term, and the corresponding noise, were replaced by a coupling to a finite set of classical degrees of freedom forming the ''quantum bath'', and the parameters characterizing the
evolution of the quantum bath were fitted to obtain the correct energy distribution for a quantum harmonic oscillator. In Ref.  \cite{dammak-prl2009}, a slightly different approach was used. The friction term $\gamma$ was maintained Markovian, and the noise was replaced by a colored noise with power spectrum $\tilde{\Theta}(\omega)= \hbar\omega \left(
\frac{1}{2} + (\exp(\hbar\omega/k_BT)-1)^{-1}
\right)$. In this manner, the correct energy distribution for the harmonic oscillator was directly obtained, and a very satisfactory agreement with low-temperature experimental data was demonstrated. The only drawback of the approach proposed in Ref. \cite{dammak-prl2009} is practical and concerns the colored noise generation. The authors used a method which is classical in the physics community \cite{billah-pra1990,lu-pre2005,rodriguez-arXiv}, but which requires to perform an inverse Fourier transform of the power spectrum of the noise over the entire time duration of the trajectory. The complete noise signal must therefore be generated and stored in the computer memory before starting the simulation. The corresponding memory requirement, for large numbers of particles or for long trajectories, is important, making the implementation of the method rather cumbersome. Our contribution in the present paper is to show how this problem can be circumvented by using a synthesis of the noise employed in the signal processing community \cite{oppenheim-book,rodriguez-arXiv}, based on an appropriate linear filter applied to a white noise, thus making the quantum thermal bath method much easier to implement in practice.

The manuscript is organized as follows. Section \ref{sec:1} recalls the basis of the method and discusses some physical issues
related to convergence. Section \ref{sec:2} describes the synthesis of the noise using a linear filter. Applications to two selected examples are considered in Section \ref{sec:3}.

\section{The quantum thermal bath method}
\label{sec:1}

The principle of the quantum thermal bath is readily illustrated on the simple case of an harmonic oscillator.
Using a unit mass and frequency $\omega_0$, the equations of motion are
\begin{equation}
\dot{X}=V
\end{equation}
\begin{equation}
\dot{V}=-\omega_0^2 X -\gamma V + \sqrt{2 \gamma}~\theta(t)
\end{equation}
where $X$ and $V$ are the position and velocity of the oscillator, $\gamma$ the friction and
$\theta(t)$ the external noise. The Fourier components are easily obtained as
\begin{equation}
X(\omega) = \sqrt{2 \gamma} \frac{\theta(\omega)}{\omega_0^2-\omega^2 +i\omega \gamma}
\label{eq:Xomega}
\end{equation}
and
\begin{equation}
V(\omega) = \sqrt{2 \gamma} \frac{i \omega \theta(\omega)}{\omega_0^2-\omega^2 +i\omega \gamma}.
\end{equation}
And finally the average energy of the oscillator is given by:
\begin{equation}
E= \int \frac{d\omega}{2\pi} \left( \frac{1}{2} \omega_0^2 \vert X(\omega) \vert^2
+\frac{1}{2}  \vert V(\omega) \vert^2 \right) = \int \frac{d\omega}{2\pi}
\gamma \frac{\omega^2+\omega_0^2}{(\omega^2-\omega_0^2)^2 +\omega^2\gamma^2} \tilde{\Theta}(\omega)
\label{energy}
\end{equation}
where $\tilde{\Theta}(\omega)$ is the power spectral density (PSD) of the noise. In the following, we will use capital letters and
a $\tilde{}$ to denote
spectral densities associated with a given random noise.

For a constant $\tilde{\Theta}(\omega)$ (white noise), integration of Eq. \ref{energy} shows that the energy is constant and independent of $\gamma$. If moreover $\tilde{\Theta}(\omega) = k_BT$ as required by the classical fluctuation-dissipation theorem, the energy is just equal to the thermal energy $k_BT$.

For a colored noise with a finite frequency support $[-\Omega_{max},\Omega_{max}]$, Eq. \ref{energy}
yields, provided $\Omega_{max} > \omega_0+\gamma$,
  $E\simeq \tilde{\Theta}(\omega_0)$ with an accuracy that depends on $\gamma$ (typically the actual value of the energy will average $\tilde{\Theta}(\omega)$
 in the frequency range $[\omega_0-\gamma/2,\omega_0+\gamma/2]$). Hence, by choosing
 \begin{equation}
\tilde{\Theta}(\omega) = \hbar |\omega| \Big( \frac{1}{2} + \frac{1}{\exp(\hbar |\omega| / k_B T) - 1} \Big)
\label{eq:Theta}
\end{equation}
and a value of $\gamma$ small compared to $\omega_0$, one ensures that the energy of the oscillator is given by the
Bose-Einstein formula including the zero point energy. The reasoning is readily extended to a set of linearly coupled harmonic oscillators
(since the friction and covariance of the noise are diagonal matrices). In this case, each eigenmode will have an energy equal to that of an harmonic oscillator with the corresponding frequency.

A slight word of caution is in order here. The spectral density given in Eq. \ref{eq:Theta} does not have a finite support, and moreover it diverges proportionally to $\omega$  for high frequencies. The integrand in Eq. \ref{energy} thus decays as $1/\omega$ for large $\omega$ and the integral diverges at high frequency for any finite value of $\gamma$.
Clearly this divergence is associated with the fact that the quantum thermal bath includes fluctuations at arbitrary high frequencies, which
are picked up by the broadening of the oscillator response function due to the damping $\gamma$. These high frequency fluctuations, however, are not expected to be present in any real system: in a crystal, the actual bath of oscillators that a given oscillator couples with
is physically limited to frequencies of the order of the Debye frequency. No higher frequency is present in the system. Hence, it appears reasonable to modify Eq. \ref{eq:Theta} by introducing an upper cutoff frequency $\Omega_{max}$ of the order of a few times the highest
frequencies observed in the system. We will see in the following that such a cutoff frequency appears most naturally in a numerical implementation of the colored noise. The divergent contribution to the energy in Eq. \ref{energy} then behaves as $\gamma \ln(\Omega_{max})$, and can be made arbitrarily small by choosing a small enough value of $\gamma$, once $\Omega_{max}$ has been specified. In general, and for a given cutoff frequency, the independence
of the results on the damping parameter $\gamma$ should be checked to ensure that the relevant spectrum of frequencies is properly thermostated.

On the other hand, it is also clear from Eq. \ref{energy} that, for a fixed cutoff $\Omega_{max}$, the result will be a decreasing function of the
damping $\gamma$. In practice, $\gamma$ should be kept small enough (compared to $\omega_0$) that the integrand actually behaves like a $\delta$ function,
while the equilibration time remains reasonable. We will see in the following that values of $\gamma/\omega_0 \sim 10^{-2}$ provide a good compromise between these various requirements.

\section{Noise synthesis}
\label{sec:2}

The numerical implementation of the quantum bath as described above reduces to generating a colored noise with a prescribed PSD, $\tilde{\Theta}(\omega)$.
 We use here a classical method of signal processing based on filtering a white noise signal \cite{oppenheim-book}. For a continuous noise, we introduce the filter
\begin{equation}
\tilde{H}(\omega) = \sqrt{\tilde{\Theta}(\omega)}
\end{equation}
  and its inverse Fourier transform $H(t)$ ($\tilde{\Theta}$ is given in Eq. \ref{eq:Theta}).
  The noise $\theta(t)$ is then obtained by convoluting $H(t)$ with a random white noise $r(t)$ with PSD $\tilde{R}(\omega) = 1$. We have:

\begin{equation}
\theta(t) = \int_{-\infty}^{\infty} H(s) r(t-s) ds
\end{equation}

The spectral density of the resulting noise is then $|\tilde{H}(\omega)|^2 \tilde{R}(\omega) = \tilde{\Theta}(\omega)$, which is the target PSD. This scheme can be implemented in a discrete algorithm suitable for computer calculation. The filter $\tilde{H}(\omega)$ is first discretized in $2 N_f$ values with steps $\delta \omega$ over an interval $[-\Omega_{max},\Omega_{max}[$:

\begin{equation}
\tilde{H}_k = \tilde{H}(k \delta \omega), k=-N_f ... N_f-1.
\label{eq:Hk}
\end{equation}

The cutoff frequency $\Omega_{max}$ can be chosen on physical grounds as discussed above, and will typically be taken to be a few times the
maximum  eigenmode frequency of the system. A discrete Fourier transform yields
 $H(t)$ over the interval $[-\pi/\delta \omega,\pi / \delta \omega[$ with a timestep $h = \pi/ \Omega_{max}$. $\tilde{H}(\omega)$ being even and real, so is $H(t)$:

\begin{equation}
H_n = \frac{1}{2 N_f} \sum_{k=-N_f}^{N_f-1} \tilde{H}_k \cos \big( \frac{\pi}{N_f} k n \big).
\label{eq:Hn}
\end{equation}

The convolution is then performed discretely in the range $[-\pi/\delta \omega,\pi / \delta \omega[$, yielding a discrete colored noise with a time step $h$:

\begin{equation}
\theta_n = \theta(n h) = \sum_{m=-N_f}^{N_f-1} H_m r_{n-m},
\label{eq:convolution}
\end{equation}

where $r_m$ is a white noise drawn from a gaussian or a uniform distribution with a variance  $\sqrt{h}$ in order to yield a unity PSD.

In practice, $\{H_m\}_{m=-N_f ... N_f-1}$ is first computed using Eqs. \ref{eq:Hk} and \ref{eq:Hn} and stored.
The noise generator is then initialized by drawing $2 N_f$ initial values $\{r_m\}_{m=0 ... 2 N_f-1}$ and time is initialized to $N_f h$. Then, at each subsequent time step,
  with time $nh$ and $n \geq N_f$, $r_{n-N_f}$ is discarded and a new random number $r_{n+N_f}$ is drawn and stored.
  The correlated noise is then obtained by computing the convolution in Eq. \ref{eq:convolution}. The present algorithm has a CPU computational cost close to a classical Langevin thermostat since the only difference is the convolution in Eq. \ref{eq:convolution}. The main difference is the memory storage because the present algorithm requires to store $4 N_f$ values per degree of freedom, corresponding to the filter and white noise. This
   requirement can however be easily managed and is much less than in the method used by Dammak \emph{et al} \cite{dammak-prl2009}, where an inverse Fourier transform is performed to create a signal
   over the complete duration of the simulation with a discretization that corresponds to the time step used by the MD integrator,
 i.e. a required  storage (number of time steps) $\times$ (number of degrees of freedom).

\begin{figure}[ht]
\includegraphics[width=0.9\linewidth]{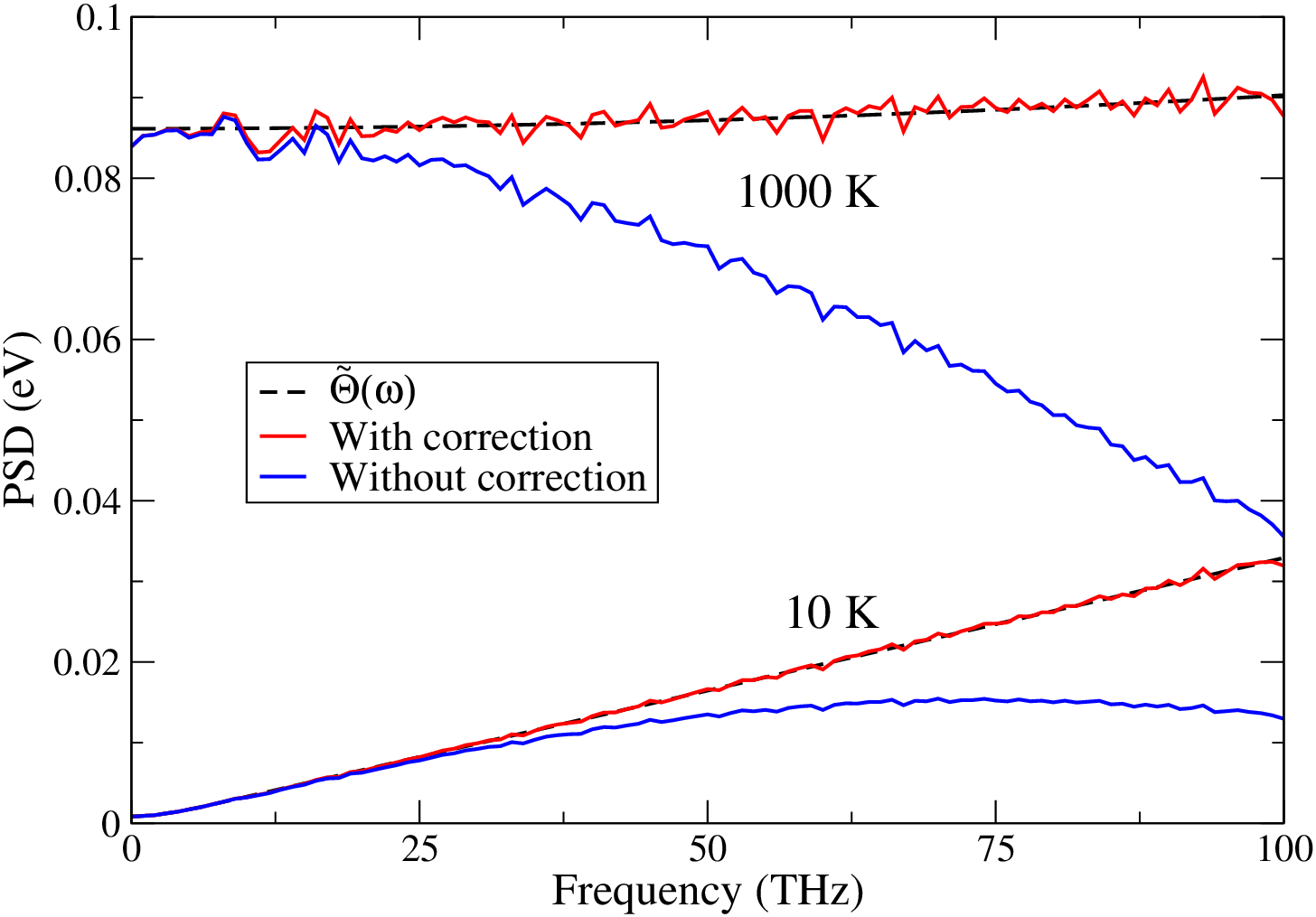}
\caption{\label{fig:NOISE}  (Color online) Noise generation at different temperatures. The dashed lines show the target
 PSD $\tilde{\Theta}$ at $T=10$ K and $T=1000$ K. The blue curves are the
 PSDs of the noise generated without correction and the red curves are after correction. 
 The parameters of the quantum thermal bath used here are: $\Omega_{max} = 100$ THz, $N_f = 100$, $M = 30$.}
\end{figure}

In principle, there is no specific reason to set the time step of the MD integrator ($\delta t$) equal to the time step of the noise generator $h$ (which corresponds to the inverse of the cutoff frequency $\Omega_{max}$). Taking again the harmonic oscillator as an example,
a reasonable time step for integration of the equation of motion is $\delta t =  0.01\omega_0^{-1}$. The cutoff frequency for the noise, on the other hand, should be physically limited to a few times $\omega_0$, let us say for concreteness $\Omega_{max}= 2\omega_0$. Hence the noise time step can be taken to be $M$ times larger than the MD time step, with $M=50$ for this specific example.
 The simplest way to manage the two time steps is to generate the noise according to the procedure outlined above, with a time step $h=M\delta t$, and to keep the noise value constant for $M$ integration  time steps $\delta t$. This amounts to generating a noise with an actual power spectrum
 \begin{equation}
\tilde{\Xi}(\Omega) = \tilde{H}(\omega)^2 C(\omega)^2,
\end{equation}
where
\begin{equation}
C(\omega) = \frac{\sin (\omega h/2)}{\omega h/2}
\end{equation}
is the Fourier transform of the square function of amplitude unity and duration $h$. Obviously the function $\tilde{\Xi}(\omega)$ is not the desired
PSD $\tilde{\Theta}(\omega)$, but this problem is easily cured by replacing $\tilde{H}(\omega)$ in Eq. \ref{eq:Hk} by a corrected spectrum
\begin{equation}
\tilde{H}_1(\omega) = \tilde{H}(\omega) C(\omega)^{-1}
\end{equation}
so that the proper spectral density is generated over the interval $[0,\Omega_{max}]$. As an illustration, Fig. \ref{fig:NOISE} compares on a specific example the PSD of the noise generated by the above procedure with or without the final correction. Before correction, the intensity of the noise is below $\tilde{\Theta}$ and decreases rapidly at high frequencies. The reason is the suppression of the high frequencies due to the noise plateaus between successive updates. On the other hand, after correction, the PSD of the noise matches very well $\tilde{\Theta}$ up to the cutoff frequency $\Omega_{max}$. The results of the quantum thermal bath are illustrated below for two simple examples.

\bigskip

\section{Examples}
\label{sec:3}

\subsection{Linear chain of oscillators}

\begin{figure}
\includegraphics[width=0.9\linewidth]{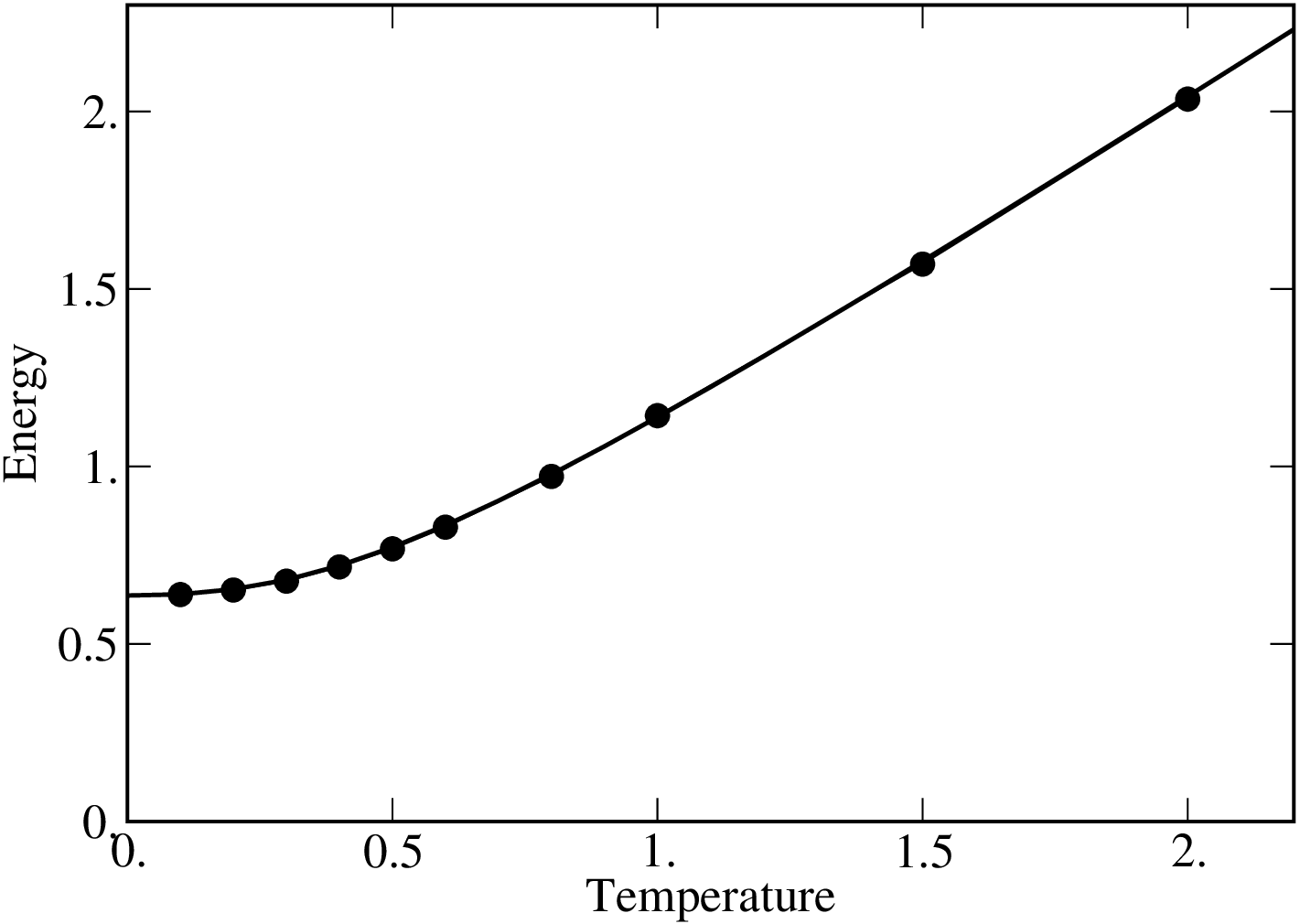}
\caption{ (Color online) Reduced energy  (in units of $\hbar \omega_0$) as a function of reduced
temperature $\hbar \omega_0/k_B$, for
 a chain of 50 oscillators. The dots are the result from a numerical calculation using the implementation of the quantum bath described in the text, with $\Omega_{max}=2\omega_0$, $\gamma=0.02\omega_0$,
and $\delta \omega =0.02 \omega_0$, i.e. $N_f = 50$. The full line is the exact energy.}
\label{fig1}
\end{figure}

We consider first the very simple example of a linear chain of oscillators
coupled by Hookean springs between nearest neighbors, with periodic boundary conditions. The frequency of an individual spring is $\omega_0$.
In Fig. \ref{fig1} the energy obtained from a MD run using a thermostat with $\Omega_{max}= 2\omega_0$ and
 $\gamma= \omega_0 /50$
is compared to the exact result for this system.

\begin{figure}
\includegraphics[width=0.9\linewidth]{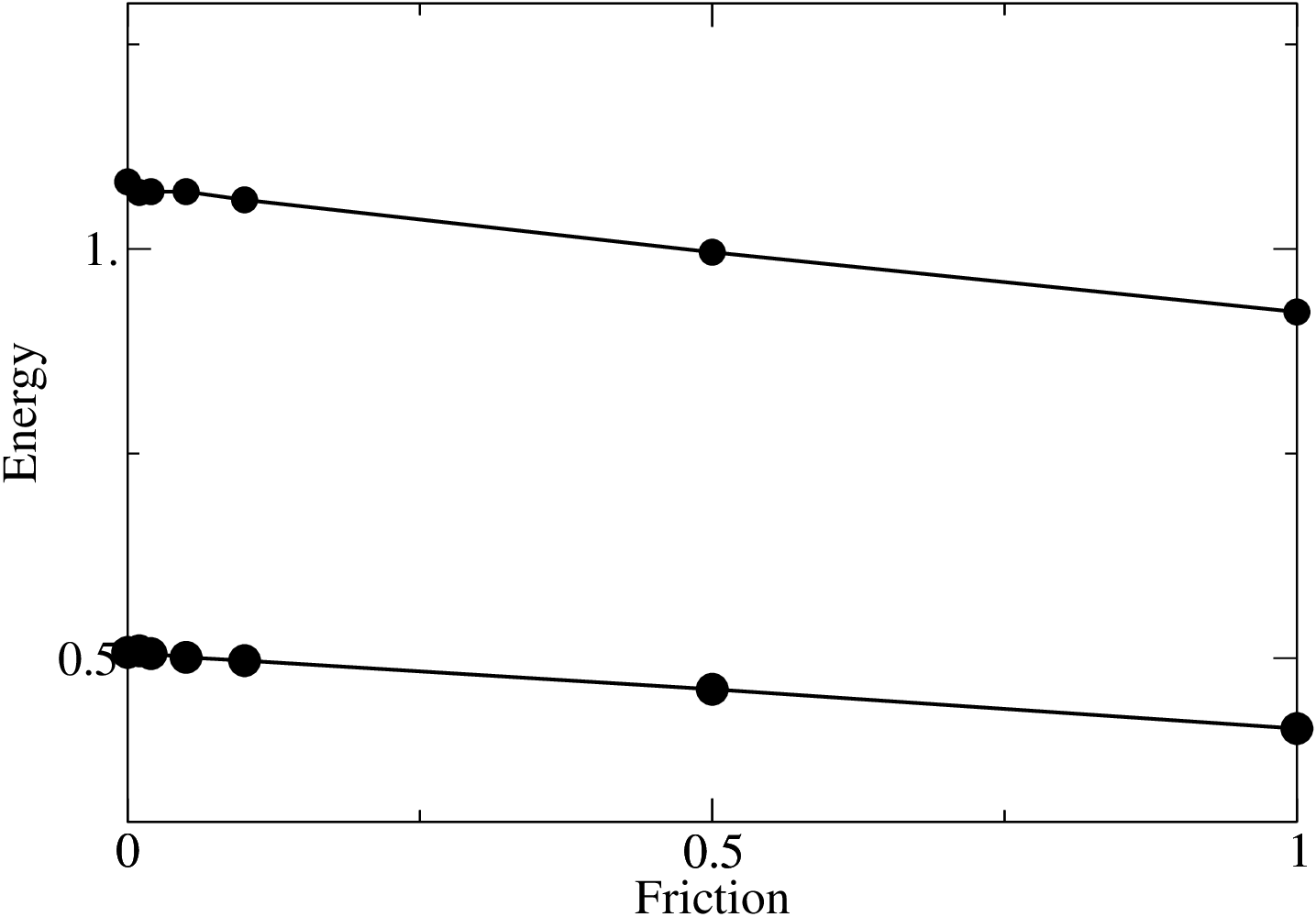}
\caption{ (Color online) Reduced energy  (in units of $\hbar \omega_0$) for  reduced
temperatures $\hbar \omega_0/k_B=0.2$ and $\hbar \omega_0/k_B=1.0$ for
 a single oscillator, as a function of the friction coefficient  $\gamma/\omega_0$. The point at $\gamma=0$
  is the exact energy.}
\label{fig2}
\end{figure}

In Fig. \ref{fig2}, we show the influence of the friction parameter $\gamma$. As expected, a too large $\gamma$ results in a broadening of the oscillator response function which does not pick up the correct  range of frequency spectrum. As the spectrum has been cut off to a rather low frequency here, the result is an artificial decrease of the energy. However, a value of $\gamma$ of the order of $\omega_0/50$ provides both a good approximation to the spectrum and reasonable equilibration time.

\subsection{Aluminium crystals}

\begin{figure}
\includegraphics[width=0.9\linewidth]{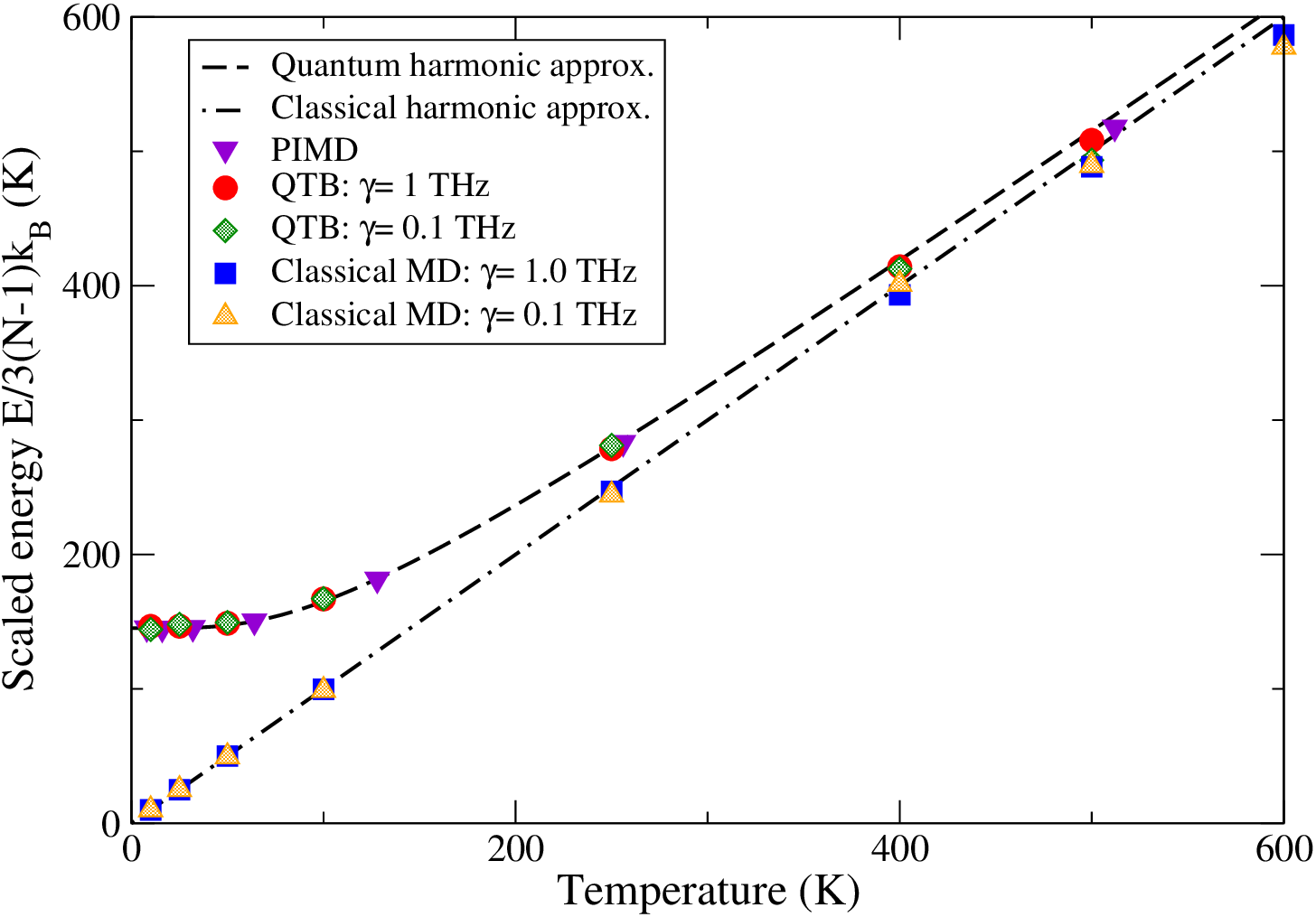}
\caption{\label{fig:ENERGY_Al} (Color online) Average energy per atom in an Aluminum crystal as a function of temperature. Different frictions $\gamma$ are used, averages over $5 \times 10^4$ MD steps after $5 \times 10^4$ steps of equilibration. Parameters of the quantum thermal bath: $\omega_{max} = 101.34$ THz, $N_f = 100$, $\delta t = 1$ fs, $h = 31 \delta t$. Results are compared to quantum and classical harmonic approximations computed from the phonon spectrum, as well as PIMD simulations with a number $P$ of replica satisfying $PT = 4096$ K.}
\end{figure}

We model an aluminum perfect crystal using the many-body embedded atom method potential developed by Ercolessi and Adams \cite{ercolessi-el1994}. The simulation cell contains a perfect periodic crystal made of $6\times6\times6$ repeating cells with crystallographic orientation $X=[100]$, $Y=[010]$ and $Z=[001]$. The cell comprises N=864 atoms. We compute the evolution of the crystal energy and Debye-Waller factor as a function of temperature using the quantum thermal bath thermostat. In both cases, the results of the MD simulations are compared to classical mechanics calculations performed using a classical Langevin thermostat, to path-integral molecular dynamics (PIMD) calculations (see Ref \cite{muser_jcp2001} for details) and to quantum and classical harmonic approximations based on the phonon frequency distribution of the crystal, $\{\omega_i\}_{i=0;3N-1}$. The latter is computed after diagonalization of the Hessian matrix of the crystal. In the harmonic approximation, the quantum average energy is given by:

\be
\langle E_{harmo} \rangle = \sum_{i=0}^{3N-4} \tilde{\Theta}(\omega_i),
\ee
where the 3 translational modes have been omitted and $\tilde{\Theta}$ is the harmonic oscillator energy given in Eq. \ref{eq:Theta}. The classical harmonic approximation is obtained by replacing $\tilde{\Theta}$ by $k_BT$. The comparison is shown in Fig. \ref{fig:ENERGY_Al}. The quantum thermal bath allows to recover very well the quantum harmonic approximation at both frictions $\gamma=0.1$ THz and 1 THz and the figure shows the convergence to the classical regime and the gradual slight departure from the harmonic approximation at high temperatures. The highest frequency in the crystal being $\omega_{max}=59.7$ THz, we conclude again that a ratio $\gamma/\omega_{max}$ in the range $1/50$ to $1/500$ is sufficient to ensure the convergence of the energy. In this example, $N_f = 100$, meaning that the memory requirement is about 100 times larger than would be the case for a classical molecular dynamics simulation, independently of the duration of the simulation.  The present algorithm is therefore much less demanding than the previously proposed implementation \cite{dammak-prl2009}, in which memory storage was proportional to the duration of the simulation.

The second example is the Debye-Waller factor computed here as the atomic mean-square displacement $\langle u^2 \rangle = \frac{1}{N}\sum_{i=1}^N \langle (\vec{r}_i(t)-\vec{r}_i(0))^2 \rangle$. In the harmonic approximation, the Debye-Waller factor is expressed as:
\be
\langle u^2 \rangle = \frac{1}{N-1} \sum_{i=0}^{3N-4} \frac{\hbar}{2M \omega_i}\coth \Big(\frac{\hbar \omega_i}{2 k_B T} \Big) = \frac{1}{N-1} \sum_{i=0}^{3N-4} \frac{\tilde{\Theta}(\omega_i)}{M \omega_i^2}
\ee

\begin{figure}
\includegraphics[width=0.9\linewidth]{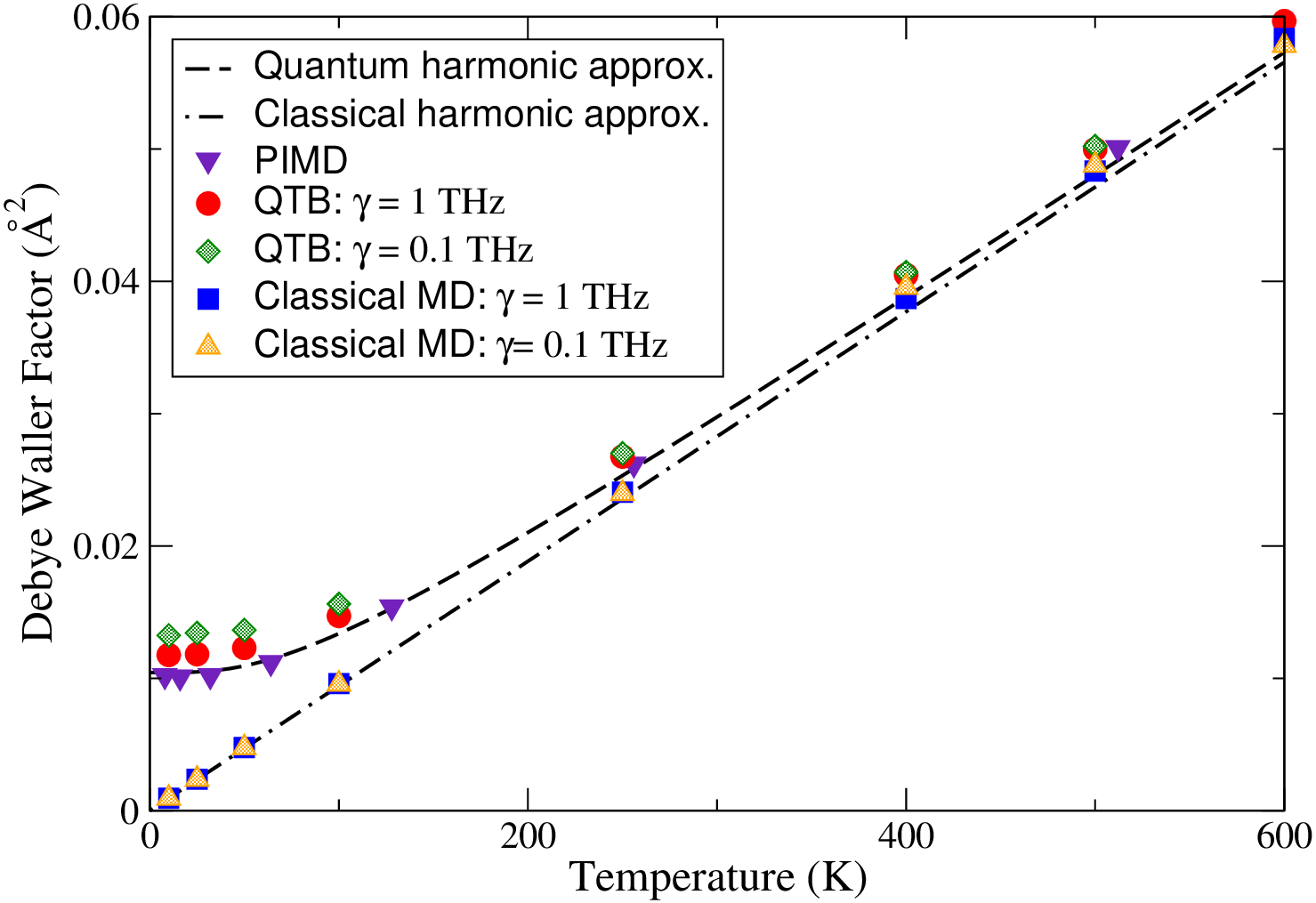}
\caption{\label{fig:DWF} (Color online) Debye-Waller factor, or average mean square displacement, in an Aluminum crystal as a function of temperature. The same parameters are used here as in Fig. \ref{fig:ENERGY_Al}.}
\end{figure}

The result is shown in Fig. \ref{fig:DWF}, where again a good agreement is seen between the QTB MD simulations and the harmonic approximation at low temperatures. It should be noted however that, in contrast with the PIMD results, the QTB simulations tend to slightly overestimate the harmonic approximation, a feature that will be discussed in section \ref{sec:5} below.

\section{Some words of caution}
\label{sec:5}

In this section, we point out some  differences between this method and the
use of a classical Langevin thermostat. With a classical thermostat, the equations of motion can be transformed
into a Fokker-Planck equation for the probability distribution of the positions and momenta. This Fokker-Planck equation admits as a stationary
solution the classical Gibbs-Boltzmann  distribution, independently of the manner in which each specific coordinate is coupled to  the thermostat.
In particular, with a set of weakly coupled (eg. trough a small anharmonic term in the energy) harmonic oscillators, the coupling of a finite number
of oscillators to the thermostat is sufficient to ensure the thermalization of the whole system at the same temperature $T$.

\begin{figure}
\includegraphics[width=0.9\linewidth]{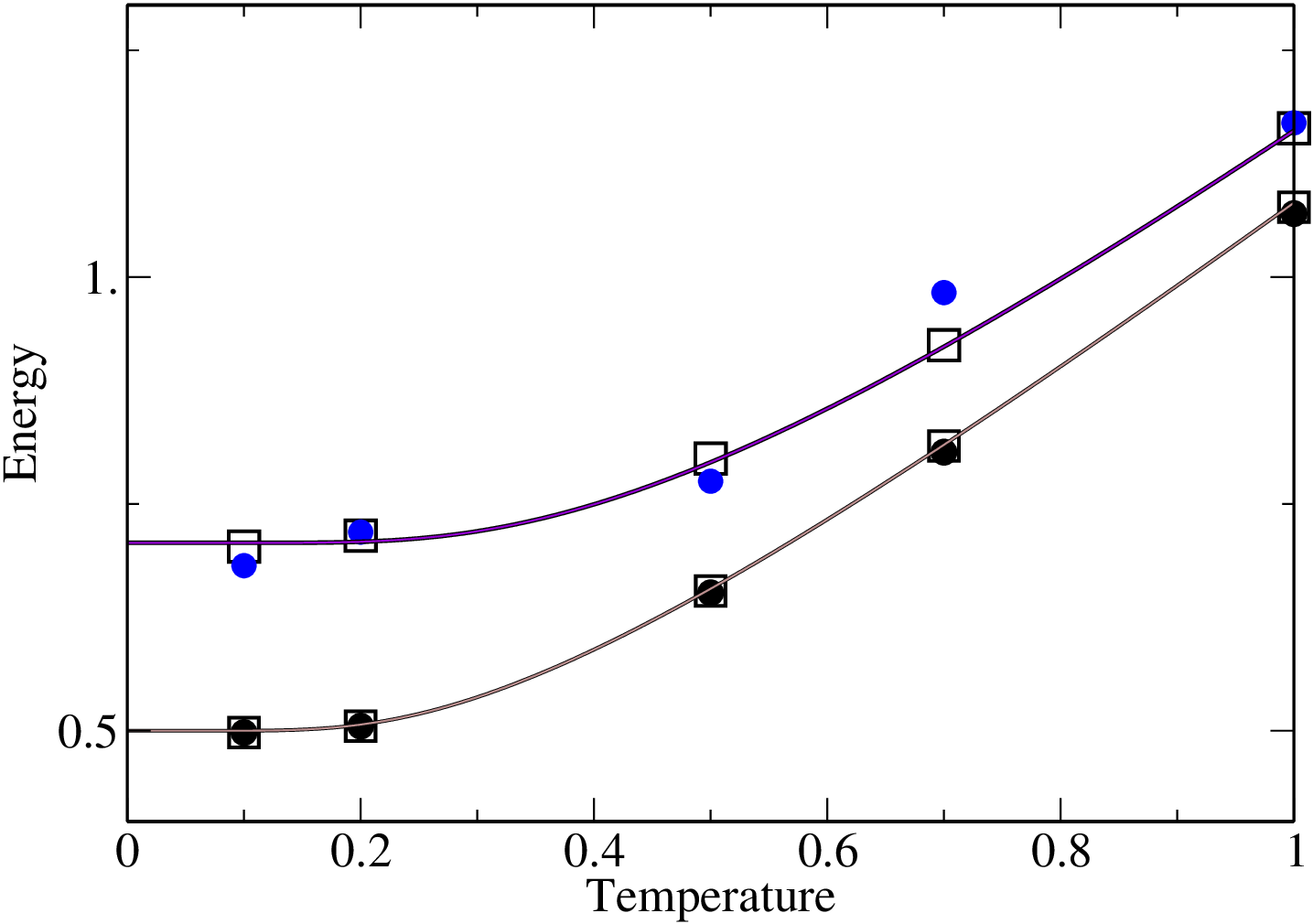}
\caption{\label{fig:coupled} (Color online) Reduced energies (measured in units of $\hbar\omega_1$) of two coupled quantum oscillators with frequencies
$\omega_1$ and $\omega_2$, as a function of the reduced temperature (measured in units of $\hbar\omega_1/k_B$). The oscillator with the
smaller frequency  ($\omega_1$) is directly coupled to a thermal bath with the same  parameters  as used in figure
\ref{fig1}. The second oscillator has a frequency $\omega_2= \sqrt{2} \omega_1$ and is not coupled to the thermal bath.
The coupling between the two oscillators is ensured by a linear (squares) or quartic (dots) term $H_c$ in the energy (see text). In both cases the strength
of the coupling is $\epsilon=0.05$ in reduced units. The points correspond to simulation data, the lines are the exact energies for decoupled oscillators.
}
\end{figure}

In the case of colored noise, the transformation to a Fokker Planck equation is not possible, and there is to our knowledge no general expression
available for the stationary phase space distribution. The quantum thermal bath presented above is coupled to all degrees of freedom,
and ensures that each independent harmonic mode acquires the correct quantum energy, independently of the coupling between modes. One may however
question the influence of such a coupling, and in particular the manner in which a ''quantum'' mode would transfer energy to other modes
that are not directly coupled to the quantum bath.  We have performed an exploratory numerical study  of this situation
by considering two weakly coupled oscillators, $X_1$ and $X_2$, with frequencies $\omega_1$ and $\omega_2$ and identical masses, and coupled
trough a weak coupling Hamiltonian $H_c(X_1,X_2)$. The first oscillator is coupled to the quantum thermal bath, while the second one evolves according to the classical
equations of motion, $M\ddot{X_2}= -M\omega_2^2 X_2 - \partial_{X_2} H_c$. We have considered the case of a linear coupling $H_c= \epsilon X_1 X_2$,
and the case of a nonlinear coupling  $H_c= \epsilon X_1^2 X_2^2 /2$. In both cases, we find (see figure \ref{fig:coupled}) that both oscillators equilibrate,
 within numerical accuracy,  at the desired energy. The equilibration time in the case of a nonlinear coupling is however extremely long,
 and the error  on the energy of the second oscillator remains significant
even after $10^7$ periods of oscillation.

In the more practical case of a weakly anharmonic system with a large number of modes, such as a crystal at low temperatures, a related question arises.
Here each mode is coupled to the quantum bath with a coupling strength $\gamma$. However, the equation of motion is also affected by the coupling to
other modes through the anharmonic terms of the energy. Classically, such anharmonic couplings give rise to a finite lifetime of the mode, characterized by an
effective damping $\Gamma_{anh}$. In order for the quantum thermal bath to operate properly, it is desirable that the coupling strength
$\gamma$ is stronger than this anharmonic coupling $\Gamma_{anh}$. On the other hand, as discussed previously, a large $\gamma$ also leads to inaccurate results for a
single mode.  One therefore has to find a reasonable compromise, in which the coupling is strong enough so that the mode can be thermalized
within its lifetime $\Gamma_{anh}^{-1}$, but weak enough to keep the thermalisation accurate. In fact, the slight deviation observed for the Debye-Waller
factor in Fig. \ref{fig:DWF} can be ascribed to the failure in finding a completely satisfactory compromise. By computing the lifetime of the short wavelength
phonons using a classical simulation at a temperature that corresponds to the quantum zero point energy, we find indeed that the order of magnitude of
the lifetime is of a few picoseconds for the present interatomic potential. This implies that the friction $\gamma$ should be at least in the Terahertz range, which indeed yields the best results for the
Debye-Waller factor. It is found that smaller values of $\gamma$ yield a repartition of energy among the normal modes of the crystal that tends to a classical distribution, independent of the mode frequency.

\section{Conclusions and perspectives}
\label{sec:6}

In this manuscript, we have presented an ''on the fly'' implementation of a quantum thermal bath for molecular dynamics simulations, suitable for long calculations in large  systems. Two simple test cases were considered as validations. The present algorithm is easily portable with limited memory requirements, and therefore opens the possibility of studying complex systems at a relatively modest computational cost.

Direct applications of interest in this context include, for static properties, all cases in which the use of the harmonic
approximation requires a numerically demanding matrix diagonalisation. This is the case in particular of amorphous or nanocrystalline
 systems at low temperatures.
Using the quantum thermal bath allows one to access directly thermodynamic quantities from a simple molecular dynamics calculation.

A more involved potential application, which may require further developments, is the investigation of thermal transport properties in
nanostructures and nanostructured materials, for which the use of classical approximations based on phonon spectra and lifetimes is
not practical. The study presented in section \ref{sec:5} suggests that modes that are only indirectly coupled to the thermal bath also equilibrate correctly at their quantum energy. Therefore it should be possible, as is done in classical heat transfer simulations, to equilibrate a temperature profile by connecting a sample to two quantum thermal baths operating  at different temperatures and in different regions of space.

Finally,  the fundamental properties of the quantum thermal bath, and in particular of the phase space distribution created by the colored noise
in a system which is not strictly harmonic, should also be investigated in the future.

\begin{acknowledgements}
We thank Hichem Dammak, Yann Chalopin, and Jean-Jacques Greffet for useful correspondence. We are particularly grateful to
Olivier Michel (Gipsa-Lab, Grenoble) who suggested the method used for the synthesis of the noise. The work was initiated during a stay at the Kavli Institute for Theoretical Physics, Santa Barbara. The authors are members of  Institut Universitaire de France.
\end{acknowledgements}

\bibliographystyle{unsrt}
\bibliography{QTB}   

%
%

\end{document}